\documentstyle[preprint,aps,epsfig]{revtex}
\global\firstfigfalse
 
\begin{document}

\draft
\preprint{$
\begin{array}{l}
\mbox{AZPH-TH/96-11}\\[-3mm]
\mbox{LBL-38582}
\end{array}
$}
\title{Dilepton and Photon Productions from a \\ Coherent Pion Oscillation}
\author{Zheng Huang\footnote{huang@physics.arizona.edu} and
Xin-Nian Wang\footnote{xnwang@nsdssd.lbl.gov}}

\address{{$^*$Department of Physics,
\baselineskip=10pt University of Arizona, Tucson, AZ 85721, USA}\\
{$^\dagger$Nuclear Science Division, MS 70A-3307 \\
\baselineskip=-8pt Lawrence Berkeley National Laboratory, 
Berkeley, CA 94720, USA}}

\date{April 10, 1996}
\maketitle
\begin{abstract}
Since the electromagnetic current for a pion system
coincides with  the third component
of the isovector current, the isospin angular oscillation of a coherent 
field can be a significant source for the electromagnetic emissions.
We study the characteristic 
dilepton and photon emissions from the classical pion field oscillation in 
the QCD vacuum.
The general analytical solution obtained in the nonlinear
$\sigma$ model is used to calculate the electromagnetic current density,
which exhibits a light-front singularity and decreases rapidly
as inverse square of the proper time due to a longitudinal expansion. 
The momentum and invariant mass spectra of the direct photon
and dilepton are found to be a sensitive
probe of the space-time evolution of the chiral condensate field.
\end{abstract}

\newpage

Recently, there has been some interest in the formation of a disoriented 
chiral condensate (DCC) in high energy hadronic and 
heavy-ion collisions \cite{dcc}. 
Preliminary studies have suggested that there may be a coherent pion 
field oscillation following a nonequilibrium second order phase transition
which may lead to characteristic soft pion production \cite{muller,num}. 
According to a
``Baked Alaska'' scenario suggested by Bjorken, Kowalski and Taylor \cite{dcc},
 the collision debris in a 
high energy central event form a ``hot'' source
on the light front and move outward while
 emitting quasi-goldstone bosons in the cool interior of light cone.   
These pions are not assumed to reach a thermal equilibrium, rather, they 
behave collectively and a coherent field description is more relevant. The
space-time evolution of the condensate 
field is mainly determined by the classical equation
of motion for a low energy effective theory such as a 
linear or nonlinear $\sigma$ model.
Eventually, the ``disoriented vacuum'' relaxes back to normal vacuum by
emitting physical pions. It is, however, difficult to obtain 
the information on its space-time evolution
from the final hadron spectrum. 

In contrast, the electromagnetic signature is known to be an ideal probe of 
a dense hadronic matter \cite{rev} 
owing to the fact that it escapes the strong 
interaction region once produced without further final state interactions
and thus carries the information on the 
 early dynamical evolution. In this Letter, we shall
study the dilepton and direct photon production from the classical
pion field in the context of a disoriented chiral condensate (DCC)
 or most generally a
nonequilibrium pion cloud. We shall develop a general formalism
for the dilepton and photon emissions
 in the presence of a classical electromagnetic
current. As an example, we calculate the dilepton and photon
spectra in an analytical solvable model where a general class of solutions 
are obtained by Blaizot and Krzywicki \cite{blaiz} and by Huang and Suzuki
\cite{suzuki}. 
Although the model is a simplified version of a more realistic situation, 
it captures the essential features of soft pion production such as the
light cone singularity, the boost invariance and the longitudinal expansion.
The electromagnetic spectra are found to be 
sensitive to the dynamical evolution
of the classical pion field. 
Since the electromagnetic current coincides with the third component
of the isovector current, the isospin angular oscillation of the condensate
field can be a significant source for the electromagnetic emission in
the low mass (transverse momentum) region.  

The pionic part of electromagnetic interactions is introduced by replacing
(we neglect the anomalous electromagnetic coupling)
$\partial_\mu \mbox{\boldmath $\pi$}$ by 
$\partial_\mu \mbox{\boldmath $\pi$}+
\frac{ie}{\sqrt{2}}[\mbox{\boldmath $\pi$},Q]
{\cal A}_\mu$ in the kinetic part of lagrangian
\begin{eqnarray}
{\cal L}^{\rm em}(x)=ie[\pi^-(x)\partial_\mu\pi^+(x)
-\pi^+(x)\partial_\mu\pi^-(x)]
{\cal A}^\mu (x)=-eJ^{\rm cl}_\mu (x) {\cal A}^\mu (x)\; ,
\end{eqnarray}
where ${\cal A}_\mu$ is the photon field and $J^{\rm cl}_\mu $ is the third
component of the classical isospin current $J^{\rm cl}_\mu (x)=
(\mbox{\boldmath $\pi$}(x)\times
\partial_\mu\mbox{\boldmath $\pi$}(x))_3$. To $O(e^2)$, the matrix element
for the dilepton emission from the classical field
 \cite{mcle,ruus,weld,huang} is
\begin{eqnarray}
{\cal M}=\langle \ell^+\ell^-|S^{(2)}|0\rangle =e^2\bar{u}(p_1)\gamma_\nu
\upsilon (p_2)D^{\nu\mu}(p_1+p_2)J^{\rm cl}_\mu (p_1+p_2)\; ,
\end{eqnarray}
where $J^{\rm cl}_\mu (p_1+p_2)$ is the Fourier transformation of 
$J^{\rm cl}_\mu (x)$, ${u}$ and $\upsilon$ are the Dirac spinors 
for the electron and positron, $D^{\nu\mu}$ is the photon propagator. 
The number of dileptons produced per final state phase space is 
\begin{eqnarray}
dN_{\ell^+\ell^-} & = &\sum_{\rm spin}|{\cal M}|^2\frac{d^3p_1}{(2\pi)^32E_1}
\frac{d^3p_2}{(2\pi)^32E_2}\nonumber \\
& = & \int d^4q\delta^4 (q-p_1-p_2)\sum_{\rm spin}|{\cal M}|^2
\frac{d^3p_1}{(2\pi)^32E_1}
\frac{d^3p_2}{(2\pi)^32E_2}\; .
\end{eqnarray}
The dilepton differential distribution with respect to the lepton pair
4-momentum $q$ is then
\begin{eqnarray}
\frac{dN_{\ell^+\ell^-}}{d^4q}=\frac{\alpha^2}{6\pi^3}\frac{B}{q^4}
[q^\mu q^\nu-q^2g^{\mu\nu}]J^{\rm cl}_\mu (q)J^{\rm cl *}_\nu (q)\; ,
\end{eqnarray} 
where $B=[1+2m^2_\ell /q^2][1-4m^2_\ell /q^2]^{1/2}$ and $m_\ell$ is the
lepton mass. Similarly, the direct photon momentum distribution 
is calculated 
\begin{eqnarray}
\frac{dN_\gamma}{d^3\mbox{\boldmath $q$}/\omega_q}
=-\frac{\alpha}{(2\pi )^2}J^{\rm cl\mu}(q)J^{\rm cl *}_\mu (q)\; ,
\end{eqnarray}
where $q$ stands for the photon 4-momentum. Moreover, splitting the current
into parts parallel and orthogonal to $q_\mu$:
$ J^{\rm cl}_\mu (q)=q_\mu J^{\rm para}
(q)+J_{\mu}^{\rm tr}(q)$, one finds that
only the orthogonal part contributes to the emissions.
The dependence of the electromagnetic spectra on the space-time 
history is entirely contained in the Fourier transformation of the isospin
current.
Although
 some more realistic numerical simulations on the classical pion field
dynamics exist \cite{num} and ideally one would like to 
perform the Fourier transformation on these solutions,   
we do not attempt to do it
in this Letter. Instead, we examine some 
simple scaling solutions in the 1+1 boost
invariant model which are first obtained by Blaizot and Krzywicki \cite{blaiz}.
Since the calculation is fully analytical, we hope to understand some 
qualitative features of dilepton production from the classical field.

The lagrangian for  the nonlinear $\sigma$ model is
 \begin{equation}
{\cal  L} = \frac{f_{\pi}^2}{4}
{\rm tr} \Bigl[\partial_{\mu}\Sigma^\dagger(x)
    \partial^{\mu}\Sigma(x)\Bigr]+\frac{m_\pi^2f_{\pi}^2}{4}
{\rm tr} \Bigl[\Sigma^\dagger(x)+
    \Sigma(x)\Bigr] , \nonumber
\end{equation}
where one defines the pion field by 
$\sigma + i\mbox{\boldmath $\tau\cdot\pi$} =
f_{\pi}\Sigma$ with the constraint $\sigma =\sqrt{f_{\pi}^2 -
\mbox{\boldmath $\pi$}^2}$. 
$\Sigma$ transforms like $\Sigma\rightarrow U_L\Sigma U_R^\dagger$ under
$SU(2)_L\times SU(2)_R$ rotations. 
We are interested in an idealized boost-invariant case \cite{bj83} where
the field $\mbox{\boldmath $\pi$}=\mbox{\boldmath $n$}(\tau )
\pi (\tau ,\mbox{\boldmath $x$}_\perp)$, i.e.\
the orientation $\mbox{\boldmath $n$}$ of 
the pion field is  only function  
of the proper time
$\tau$ defined by $\tau = \sqrt{t^{2}-z^{2}}$ and uniform in
spatial rapidity and transverse coordinates 
$\mbox{\boldmath $x$}_\perp$. This would correspond to 
a single DCC domain in the transverse dimension.
The transverse profile of the field amplitude 
$\pi^2$ is determined
by the shape and size of the DCC domain, which we shall take a Gaussian
form as suggested in numerical simulations \cite{num}
\begin{equation}
g(\mbox{\boldmath $x$}_\perp)=
\exp [-|\mbox{\boldmath $x$}_\perp|^2/R_D^2]\; ,
\end{equation}
 where $R_D$ is the size of a DCC domain (typically $R_D=1\sim 2$ fm
\cite{num}). Such a class of solutions are well known to predict a distinctive
neutral pion probability distribution
\begin{equation}
\frac{dP}{df}=\frac{1}{2\sqrt{f}}\; ,\label{dpdf}
\end{equation}
where $P$ is the probability of find a neutral pion fraction
$f=N_{\pi^0}/(N_{\pi^0}+N_{\pi^\pm})$. However, if 
there exists more than one DCC domain, as it is likely the case
especially in heavy-ion collisions where the transverse dimension
$R_A$ of the interaction volume is much larger than the size 
of a DCC domain $R_D$, the field 
orientation $\mbox{\boldmath $n$}$ could depend
on the transverse coordinates $\mbox{\boldmath $x$}_\perp$. 
In this case, 
the prediction (\ref{dpdf}) in general no longer holds. Such a 
possibility can only be studied in detail in numerical simulations.
In this Letter, we shall confine ourselves to the case when the single
domain dominates and only comment on how our result might
change in a multi-domain case.
 
According to the ``Baked Alaska'' scenario \cite{dcc}, 
the boundary conditions 
for the solutions are imposed such that the classical waves propagate forward
in proper time, starting at an initial time $\tau_0$. In other words,
we are interested in a retarded wave that vanishes when 
$\tau <\tau_0$ and $t< \tau_0$. Such a solution 
only exists when there is a source located on the hyperbola of equal 
$\tau =\tau_0$. The ``hot'' source can be regarded as the summation of the
 degrees of freedom other than pion modes. Away from the source located 
near the surface of the light cone, the field 
propagates freely and one should
have the (partial) conservation of (axial) isovector currents
\begin{eqnarray}
\partial_\mu \mbox{\boldmath $V$}^\mu & = &
\partial_\mu (\mbox{\boldmath $\pi$}\times \partial^\mu
 \mbox{\boldmath $\pi$})  =  0\; ,\quad\quad 
(\tau >\tau_0)\label{vec}\\
\partial_\mu \mbox{\boldmath $A$}^\mu & = &
\partial_\mu (\sigma\partial^\mu \mbox{\boldmath $\pi$}-
\mbox{\boldmath $\pi$} \partial^\mu\sigma )  =  -m_\pi^2f_\pi 
\mbox{\boldmath $\pi$} \; .\quad\quad 
(\tau >\tau_0)\label{axi}
\end{eqnarray}
For a function only of $\tau$, a partial derivative
$\partial_{\mu}f(\tau)$ is equal to $(\widetilde{x}_{\mu}/\tau)df/d\tau$
where $\widetilde{x}_{\mu}=(t,z,0,0)$.
Eq.(\ref{vec}) yields the integration
\begin{equation}
\mbox{\boldmath $\pi$}\times \partial^\mu
 \mbox{\boldmath $\pi$}=f_\pi^2\mbox{\boldmath $a$}\frac{\widetilde{x}_\mu}
{\tau^2}g(\mbox{\boldmath $x$}_\perp )\; ,
\label{solu}
\end{equation}
where $\mbox{\boldmath $a$ }$ is 
a dimensionless constant vector in isospin space.
In the chiral limit, the axial current is also conserved, which defines
another constant vector $\mbox{\boldmath $b$}$ in isospin space with
a constraint  $\mbox{\boldmath $a$}\cdot\mbox{\boldmath $b$}=0$
\cite{blaiz}
\begin{equation}
\mbox{\boldmath $\pi$} \partial^\mu \sigma -\sigma\partial_\mu
 \mbox{\boldmath $\pi$}=f_\pi^2\mbox{\boldmath $b$}
\frac{\widetilde{x}_\mu}{\tau^2}g(\mbox{\boldmath $x$}_\perp )\; .
\label{solu2}
\end{equation}
The pion field orientation vector 
$\mbox{\boldmath $n$}(\tau )$  precesses with proper time
$\tau$ around 
$\mbox{\boldmath $a$}$ in the plane perpendicular to 
 $\mbox{\boldmath $a$}$ \cite{suzuki}.

The solution (\ref{solu}) exhibits an inverse square singularity as
$\tau$ goes to zero. 
As pointed out in \cite{blaiz}, the singular oscillation near 
$\tau =0$ does not result from the neglect of pion mass: treating
the pion mass as a perturbation  does not generate qualitative
modification of the solution at small value of $\tau$.
Since the ratio $\widetilde{x}_\mu/\tau$ can be regarded as the velocity
$u_\mu$ of a comoving 
element with coordinate $x_\mu$, the current diverges outward from the origin 
indicating the existence of a source near the space-time origin.
The boundary condition is satisfied if one multiplies(\ref{solu}) by
 a step function $\theta (\tau -\tau_0)$. 
The isovector current
in the whole space-time is 
\begin{equation}
\mbox{\boldmath $V$}_\mu (\tau ,\mbox{\boldmath $x$}_\perp)
 =f_\pi^2\mbox{\boldmath $a$}\frac{\widetilde{x}_\mu}{\tau^2}\theta (\tau 
-\tau_0)g(\mbox{\boldmath $x$}_\perp )\; ,
\end{equation}
The source term is then 
calculated (note that $\partial_\mu [\widetilde{x}^\mu f(x)]
=(\partial_t,\partial_z,0,0)[\widetilde{x}^\mu f(x)]$)
\begin{equation}
\partial_\mu \mbox{\boldmath $V$}^\mu =f_\pi^2\mbox{\boldmath $a$}
\frac{1}{\tau}
\delta (\tau -\tau_0) g(\mbox{\boldmath $x$}_\perp )
\; .\label{source}
\end{equation}
It is clear that 
the pionic part of isospin (also the electromagnetic) current
is not conserved due to the existence of a source term at $\tau =\tau_0$
while the integration constants $\mbox{\boldmath $a$}$ and
$\mbox{\boldmath $b$}$ specify the orientation of 
the source in isospace or the initial condition.
 It is also straightforward to calculate the
energy density associated with the internal chiral oscillations
using the calculated classical currents $\mbox{\boldmath $V$}$
and $\mbox{\boldmath $A$}$ and the classical
solution for the pion field \cite{blaiz,suzuki}
\begin{equation}
\epsilon =\frac{2f_\pi^2( \mbox{\boldmath $a$}^2+
\mbox{\boldmath $b$}^2)}{\tau^2}\; .\label{energy}
\end{equation}
As required by the chiral symmetry, the energy density depends only 
on the combination $\mbox{\boldmath $a$}^2+
\mbox{\boldmath $b$}^2$.

The dilepton and photon emissions probe the nature of the source 
using  time-like
momenta $q=(q_0,q_{||},\mbox{\boldmath $q$}_T)$ with $q^2=M^2$ and $q^2=0$
respectively. 
The Fourier transformation of the
electromagnetic current can be readily calculated in the light cone
variables
\begin{eqnarray}
J^{\rm cl}_\mu (q) & = & f_\pi^2a_3\int d^4x e^{iqx}
\frac{\widetilde{x}_\mu}{\tau^2}
\theta (\tau -\tau_0)g(\mbox{\boldmath $x$}_\perp )
\nonumber \\
& = & -if_\pi^2a_3\frac{\widetilde{q}_\mu}{M_\perp} 
\frac{\partial}{\partial M_\perp}
\int_{\tau_0}\frac{d\tau}{\tau}\int d\eta e^{iM_\perp\tau\cosh (\eta-y)}
\int d^2\mbox{\boldmath $x$}_\perp g(\mbox{\boldmath $x$}_\perp )
e^{-i\mbox{\boldmath $q$}_T
\cdot \mbox{\boldmath $x$}_\perp} \label{four}\\
& = & -\frac{\pi^2R_D^2}{2}f_\pi^2a_3
\frac{\widetilde{q}_\mu}{M^2_\perp}\exp [-q_T^2R_D^2/4]
[J_0(M_\perp\tau_0)-i N_0(M_\perp\tau_0)]\; ,\nonumber
\end{eqnarray} 
where $\widetilde{q}_\mu =(q_0,q_{||},0,0)$, $M_\perp =\sqrt{q_0^2-q_{||}^2}$
and $J_0$ and $N_0$ are the Bessel functions. For a direct photon,
$M_\perp =q_T$.
In deriving (\ref{four}), we have performed the $\tau$-integration from
$\tau_0$ to infinity. Since the electromagnetic current density
decreases rapidly
with the proper time, the integrated current turns out to be finite and 
the emission is most effective in the early evolution
of the classical field. It is clear from
(\ref{four}) that due to the singular behavior of the Neumann function
for small argument, the  emission spectra are  sensitive to the initial
time scale $\tau_0$. The exponential factor $\exp [-q_T^2R_D^2/4]$ provides
a strong $q_T$ cutoff so that the emission is only significant in
low momentum region.

The Fourier transformation of the electromagnetic current takes the
form $ J^{\rm cl}_\mu (q)=\widetilde{q}_\mu f(M_\perp ,q_T)$ where $f$ is
a scalar function, the orthogonal part is then
\begin{equation} 
J^{\rm tr}_\mu (q)=(0,0,-\mbox{\boldmath $q$}_T)f(M_\perp ,q_T)\; ,
\end{equation}
and 
\begin{eqnarray}
(q^\mu q^\nu -q^2g^{\mu\nu})J^{\rm tr}_\mu (q)J^{\rm tr*}_ \nu (q)
& = & q_T^2M^2_\perp |f(M_\perp ,q_T)|^2 \\
J^{\rm tr\mu} (q)J^{\rm tr*}_ \mu (q) & = & -q_T^2|f(M_\perp ,q_T)|^2 \; .
\end{eqnarray}
The dilepton and photon distributions are calculated
\begin{eqnarray}
\frac{dN_{\ell^+\ell^-}}{dM_\perp^2dydM} & = &\frac{\alpha^2}{24}
(\pi R_D^2)^2Bf_\pi^4a_3^2\frac{q_T^2}{M^3M_T^2}\exp [-q_T^2R_D^2/2]
[J_0^2(M_\perp\tau_0)+ N_0^2(M_\perp\tau_0)]\; ,\label{dilep}\\
\frac{dN_\gamma}{dydq_T} & = & \frac{\alpha}{8}\pi 
(\pi R_D^2)^2f_\pi^4a_3^2
\frac{1}{q_T}\exp [-q_T^2R_D^2/2]
[J_0^2(q_T\tau_0)+ N_0^2(q_T\tau_0)]\; ,\label{photon}
\end{eqnarray}
where $q_T^2=M_\perp^2-M^2$ for the dilepton.
Both the dilepton and the photon spectra fall off exponentially for
large transverse momentum which is the
characteristic of the coherent production from a finite domain. The 
electromagnetic emission from a DCC domain is thus only important in the
low momentum region where the spectra 
increase as some inverse power law of the momentum.  

The constant $a_3$ can
be estimated using the symmetry argument. The important assumption in the
disoriented chiral condensate scenario is the equal probability of all 
internal orientations. According to Eq.\ (\ref{energy}), the energy
density due to the internal chiral oscillations should be 
equally distributed among all components of vectors 
$\mbox{\boldmath $a$}$ and $\mbox{\boldmath $b$}$. One thus has
\begin{equation}
\langle a_i^2\rangle =\langle b_i^2\rangle =\frac{1}{6}
\frac{\epsilon_0\tau_0^2}{2f_\pi^2}\quad\quad (i=1,2,3)\; .\label{a3}
\end{equation}
where $\epsilon_0$ is the initial energy density 
carried by the internal chiral oscillationat at $\tau =\tau_0$, which
may be roughly estimated assuming that it is comparable to  the thermal
energy density (in order for the DCC to have any physical significance).  
For some typical values $\epsilon_0\sim 0.5T_c^4\sim 0.1$
 GeV/fm$^3$ and $\tau_0=5$ fm/c, 
the average value of $a_3^2$ is about $4.8$.
\begin{figure}
\centerline{\epsfig{figure=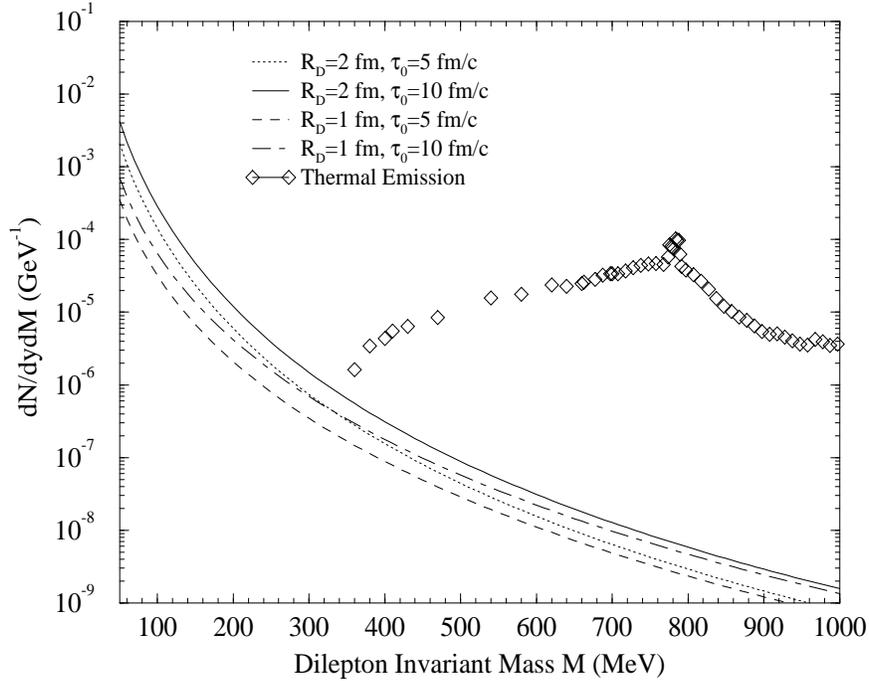,width=5in,height=4in}}
\caption{The dilepton invariant mass spectrum 
for different choices of initial time scale $\tau_0$ and the coherent
field domain size $R_D$.
The initial energy density is assumed to be  $\epsilon_0=0.1$ GeV/fm$^3$.
The typical thermal spectrum due to the 
$\pi$-$\pi$ annihilation is also plotted for a comparison} 
\end{figure}
In Fig.1 we numerically integrate (\ref{dilep}) over
$M_\perp$ and plot the dilepton invariant mass spectrum for 
different values of $\tau_0=5$, 10 fm/c and $R_D=1,2$ fm. 
The $a_3$ is calculated
from (\ref{a3}) assuming that the initial energy density
for the internal chiral oscillations is $0.1$ 
GeV/fm$^3$. 
Also plotted in Fig.1 is the typical thermal spectrum of dilepton
production mainly from $\pi$-$\pi$ annihilations, which is taken
from Ref.\cite{huang}, where we have assumed the initial
temperature of 160 MeV, the freeze-out temperature of 130 MeV,  
and the transverse size of 
system $R_A=5$ fm suitable for heavy-ion collisions. 
Most notably, there does not exist a pion mass
threshold for the dilepton production from the coherent field: the
spectrum rises even below $M=2m_\pi$. The finite pion mass plays no
roles in the conserved isovector current whose third component
 directly couples to the photon. Therefore, the coherent field
is most effective in producing lepton pairs  in the
low mass region $M<2m_\pi$. 
\begin{figure}
\centerline{\epsfig{figure=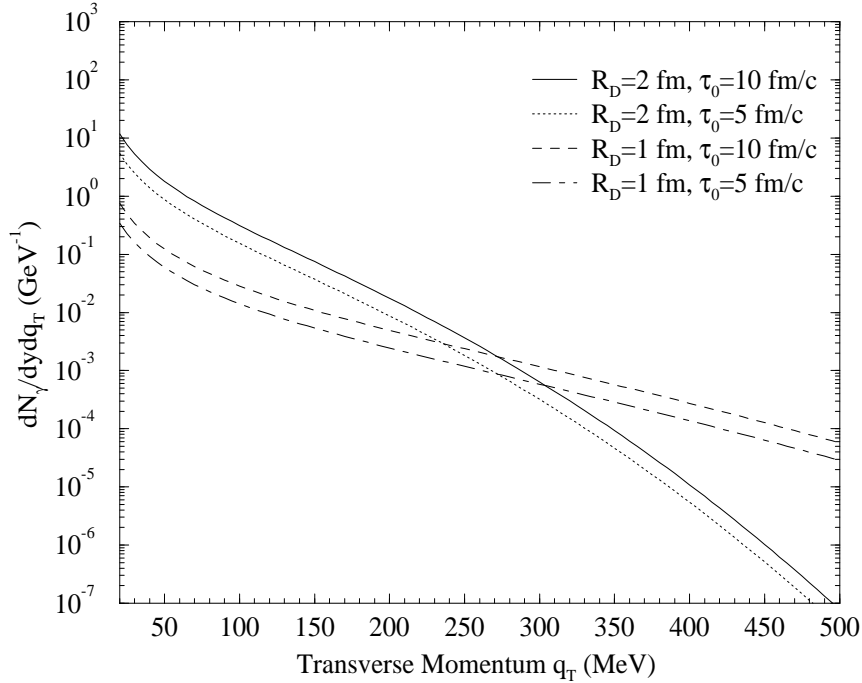,width=5in,height=4in}}
\caption{The direct photon transverse momentum spectrum 
for different choices of $\tau_0$ and $R_D$.
The initial energy density is assumed to be $\epsilon_0=0.1$ GeV/fm$^3$.} 
\end{figure}
Plotted in Fig.2 is the
direct photon transverse momentum spectrum for different
choices of $\tau_0$ and $R_D$. The outstanding feature in the 
spectrum is the sensitivity on the domain size parameter $R_D$.
The measurement of direct photon spectrum at very small transverse
momentum can thus provide a sensitive probe of the spatial extension
of the DCC domain. 

So far we have only considered a single DCC domain and its orientation
oscillation in time. 
If the multi-domain configurations dominate the DCC production, 
the dilepton and photon can be coherently produced from
different oriented regions. In particular, if there are two 
domains with opposite orientation (the molecule type), 
the singular rise of the spectra
near zero momentum will be suppressed and spectra at the moderate 
momentum will be enhanced. We plan to address these interesting
questions in future publications.
 
In conclusion, we have suggested that the study of electromagnetic
emission can provide some information on the space-time
history of a disoriented chiral condensate.  
The emissions due to the coherent isospin oscillation 
can be  an important additional source
to the electromagnetic signals  from a dense hadronic matter, especially
in low mass (momentum) region. It would be very interesting to see
whether or not the qualitative features
suggested in this paper can survive in more realistic models, and indeed
may provide some mechanism for the observed low mass dilepton enhancement
by the CERES and HELIOS collaborations at CERN SPS experiments \cite{sps}. 
 
We would like to thank Hans Thomas Elze and Ina Sarcevic  
for useful discussions.
This work was supported 
through the U.S. Department of Energy under Contracts Nos.\ 
DE-FG03-93ER40792 and DE-AC03-76SF00098.

\end{document}